\documentclass[sigconf, nonacm]{acmart}

\AtBeginDocument{%
  }
 
\begin{document}
\sloppy

\title{Survey of City-Wide Homelessness Detection Through Environmental Sensing}

\author{Julia Gersey}
\email{gersey@umich.edu}
\affiliation{%
  \institution{University of Michigan}
  \city{Ann Arbor}
  \state{Michigan}
  \country{USA}
}

\author{Rose Allegrette}
\email{rallegre@umich.edu}
\affiliation{%
  \institution{University of Michigan}
  \city{Ann Arbor}
  \state{Michigan}
  \country{USA}
}
 
\author{Joshua Lian}
\email{joshlian@umich.edu}
\affiliation{%
  \institution{University of Michigan}
  \city{Ann Arbor}
  \state{Michigan}
  \country{USA}
}

\author{Zawad Munshi}
\email{munshiz@umich.edu}
\affiliation{%
  \institution{University of Michigan}
  \city{Ann Arbor}
  \state{Michigan}
  \country{USA}
}

\author{Aarti Phatke}
\email{aphatke@umich.edu}
\affiliation{%
  \institution{University of Michigan}
  \city{Ann Arbor}
  \state{Michigan}
  \country{USA}
}

\renewcommand{\shortauthors}{Gersey et al.}

\maketitle

\section*{Abstract}

The growing homelessness crisis in the U.S. presents complex social, economic, and public health challenges, straining shelters, healthcare, and social services while limiting effective interventions. Traditional assessment methods struggle to capture its dynamic, dispersed nature, highlighting the need for scalable, data-driven detection. This survey explores computational approaches across four domains: (1) computer vision and deep learning to identify encampments and urban indicators of homelessness,
(2) air quality sensing via fixed, mobile, and crowdsourced deployments to assess environmental risks,
(3) IoT and edge computing for real-time urban monitoring, and
(4) pedestrian behavior analysis to understand mobility patterns and interactions. Despite advancements, challenges persist in computational constraints, data privacy, accurate environmental measurement, and adaptability. This survey synthesizes recent research, identifies key gaps, and highlights opportunities to enhance homelessness detection, optimize resource allocation, and improve urban planning and social support systems for equitable aid distribution and better neighborhood conditions.
\section{The Role of Environmental Sensing in Homelessness Research}

Homelessness is a multifaceted and growing crisis in the United States, driven by economic instability, rising housing costs, and insufficient social support systems. In 2023, the annual Point-in-Time (PIT) count recorded 653,104 individuals experiencing homelessness, marking a 12.1\% increase from the previous year \cite{soucy_state_nodate}. The 2024 State of Homelessness report highlights a widening gap between the increasing number of people entering homelessness and the limited resources available for intervention. Additionally, the 2020 Unsheltered Homeless research report showed that the increase in unsheltered homelessness also increased costs for municipal governments \cite{batko_unsheltered_2020}. This is due to increased municipal spending on initiatives for encampment management, sanitation upkeep, and “hostile architecture” designed to deter homeless individuals (e.g., fences, bars, rocks, spikes). This excludes indirect costs: impacts on public health and the environment. In particular, in the closing of San Jose`s largest encampment, two weeks of manual cleanup removed more than 600 tons of debris and 2,850 gallons of bio-waste \cite{batko_unsheltered_2020}. This figure does not even account for the trash and waste that entered waterways, affecting nearby communities \cite{johnson_socio-economic_2016}. As the crisis escalates, scalable, cost-effective, and data-driven approaches are essential for accurately detecting homeless locations. These methods enable more efficient resource allocation, address public health concerns, and help cities support homeless populations, while also improving neighborhood conditions and sanitation.

Traditional methods for tracking and supporting homeless populations—such as PIT counts, manual outreach, and citizen reporting—are constrained by labor-intensive processes, limited temporal coverage, and incomplete geographic reach. These limitations hinder timely, data-driven decision-making for social services and urban planning. In response, computational advancements in sensing technologies and machine learning offer an opportunity to provide continuous, scalable, and objective insights into homelessness trends and their environmental contexts.

Environmental factors significantly impact the experiences of unhoused individuals, influencing health risks, living conditions, and mobility patterns. Variables such as air quality, pollution levels, urban cleanliness, and pedestrian behavior serve as indicators for detecting encampments and informing targeted interventions. By leveraging technologies such as computer vision, air quality monitoring, Internet of Things (IoT) networks, and behavioral analytics, researchers can develop more effective methods to identify homeless populations and understand the broader urban and environmental factors contributing to homelessness.

This survey explores the role of environmental sensing in detecting homelessness, focusing on four key domains: (1) computer vision and deep learning to identify encampments and urban indicators of homelessness,
(2) air quality sensing via fixed, mobile, and crowdsourced deployments to assess environmental risks,
(3) IoT and edge computing for real-time urban monitoring, and
(4) pedestrian behavior analysis to understand mobility patterns and interactions. By reviewing recent advancements and identifying ongoing challenges, this paper highlights the potential for computational sensing methods to enhance homelessness detection, optimize resource allocation, and support more effective public health interventions. Furthermore, it discusses the limitations of existing technologies, such as computational constraints, data privacy concerns, and the adaptability of detection methods to the dynamic nature of homelessness, providing insights into future research directions.
\section{Automated Identification of Homeless Encampments Using Computer Vision and Deep Learning}

Identifying homeless encampments through computer vision and crowd-sourced data requires analyzing environmental indicators and leveraging deep learning techniques to detect makeshift shelters in urban areas \cite{alfarrarjeh_object_2023}. Detecting homelessness through computer vision and crowd-sourcing is heavily influenced by the physical characteristics of an area \cite{moon_human-centered_2024,fernandez2024poster}. Various environmental features have been analyzed to identify patterns associated with homelessness, including traffic levels \cite{ke_real-time_2019}, trash accumulation \cite{brown_sheltering_2021, rad_computer_2017}, road and parking lot deterioration \cite{ricord_impact_2020, majidifard_deep_2020,harishankar2020lanet}, building quality \cite{zhong_convolutional_2019}, and the presence of graffiti \cite{ozsvath_graffiti_2022}. These indicators provide important contextual cues that aid in the automated detection of areas with high rates of homelessness.

Building on this foundation, object detection techniques have been specifically developed to identify homeless encampments, which serve as critical sites for intervention and resource allocation \cite{toma_iot_2019}. One such approach, "TentNet", utilizes satellite imagery to detect tent structures and has achieved a success rate of 73.68\%  \cite{lam_xview_2018, bhil_recent_2022}. However, relying solely on high-resolution satellite imagery presents limitations, including high costs and the requirement for an unobstructed overhead view \cite{zhong_convolutional_2019, esposito_dynamic_2016}. To address these challenges, an alternative approach involves leveraging video footage captured via mobile surveillance, such as cameras mounted on moving vehicles or preexisting surveillance infrastructure \cite{anjomshoaa_city_2018, naik_visual_2017, hwang_systematic_2023}. By extracting frames from video streams, this method provides a more cost-effective and adaptable solution, enabling detection in areas where overhead visibility is obstructed \cite{baumgartner2013tracking}. Recent advances in deep learning, particularly through deep neural networks (DNNs), have demonstrated high accuracy in image classification tasks, further enhancing the feasibility of this approach \cite{chen_visual_2022}.

Despite their effectiveness, DNNs have several limitations, including computational redundancy, difficulty in identifying objects in low-quality or blurry images, challenges in detecting objects under extreme lighting conditions, and the high computational cost associated with GPUs and TPUs. Additionally, the processing time required for complex models can be a bottleneck in real-time applications \cite{sood_significance_2021}. Another significant limitation of this approach is its reliance on tent detection, which does not account for the full diversity of homelessness. Many individuals experiencing homelessness do not reside in tents; instead, they construct makeshift shelters or sleep directly on the ground or in sleeping bags \cite{nettleton_sleeping_2012, xu_understanding_2019}. These variations in living conditions present additional challenges for automated detection systems, necessitating further research into more comprehensive and adaptable detection methodologies.
\section{Air Quality Sensing as an Indirect Factor in Homelessness Analysis}

The Clean Air Act requires the regulation of hazardous air pollutants and has been modified to regulate 188 such pollutants. The EPA outlines primary sources to include human-made mobile sources, stationary sources, and natural sources of origination \cite{us_epa_initial_2015, us_epa_hazardous_2015}. Each pollutant poses various harms to humans through ingestion, exposure, and accumulation. As a result, air quality monitoring has become a key tool for understanding urban pollution exposure, climate effects, and public health risks. Research reflects this interest, showing that high levels of particulate matter (PM$_{2.5}$ and PM$_{10}$), nitrogen dioxide (NO$_2$), and ozone (O$_3$) are commonly found in high-traffic urban areas and industrial zones. These areas often overlap with locations where many unhoused individuals seek shelter due to their proximity to infrastructure, public spaces, or temporary resources. Thus, air quality sensing can serve as an indirect but valuable indicator for understanding the spatial distribution of homelessness and the environmental conditions affecting unhoused populations.

While broad-scale air pollution monitoring exists, traditional sensing networks often lack the granularity needed to capture the environmental conditions in specific areas where homeless populations reside. To address this, researchers have developed fixed, mobile, wearable, and crowdsourced air quality sensing methods that provide high-resolution pollution mapping and can support urban planning and public health interventions targeting environmental risks for unhoused individuals.

\subsection{Fixed vs. Mobile Air Quality Sensing}

Air quality monitoring traditionally relies on fixed sensor networks. For example, the Environmental Protection Agency (EPA)'s AirNow platform is a popular site for checking current air quality conditions, as it aggregates data from the National Oceanic and Atmospheric Administration (NOAA), NASA, the Centers for Disease Control and Prevention (CDC), and various state, city, and local air quality monitoring stations \cite{noauthor_airnowgov_nodate}. These stationary networks provide high-quality, long-term air pollution data, but their limited spatial resolution prevents them from detecting pollution variability within neighborhoods or on individual streets.

To enhance coverage, researchers have deployed fine-grained sensor networks to increase spatial resolution across cities. These deployments have leveraged low-cost sensors to monitor overall air quality, pollution levels, particulate matter, and gas concentrations, capturing real-time city data \cite{moore_air_2012, cheng_aircloud_2014, jiao_community_2016, tsujita_gas_2005, moltchanov_feasibility_2015, lagerspetz_megasense_2019, gersey_pilot_2023, krupp_towards_2023}. One study found that PM$_{2.5}$ levels doubled between two sensing units less than four miles apart \cite{krupp_towards_2023}, demonstrating the need for these systems to be deployed at a fine-grained scale. However, building and maintaining a large static sensor network can be costly in terms of equipment, labor, and connectivity. To mitigate this, some projects leverage historical data to train models that estimate pollution levels in unmonitored areas \cite{cheng_maptransfer_2020}. Still, fixed sensors face notable challenges: they require ongoing maintenance, are limited in placement by power and connectivity constraints, and struggle to achieve city-wide fine-grained coverage.

Mobile air quality sensing has emerged as a complementary approach, using drones, vehicles, and wearable sensors to measure air pollution across diverse urban environments \cite{agarwal_modulo_2020, noor2024fusion, anjomshoaa_city_2018, badii_real-time_2020, banach_new_2020, dutta_common_2009, nikzad_citisense_2012, maag_w-air_2018, hasenfratz_participatory_2012, gao_mosaic_2016, shirai_toward_2016}. Low-cost mobile sensors mounted on vehicles allow researchers to capture fine-grained, real-time pollution variations near homeless encampments, underpasses, and high-risk urban areas, offering insights for both real-time monitoring and historical analysis. Work has also explored the effect of vehicle speed and driving behavior on sensor accuracy \cite{liu_delay_2017, liu_individualized_2017, xu_gotcha_2016}. Wearable and human-carried sensors enhance spatial granularity by capturing air quality data from sidewalks, alleyways, and other locations inaccessible to vehicles.

\subsection{Crowdsourced and Large-Scale Data Collection}

Beyond mobile deployments, crowdsourced sensing expands monitoring coverage by leveraging rideshare vehicles, taxi fleets, and human-carried sensors \cite{chen_asc_2019, chen_pas_2020, xu_ilocus_2020, xu_gotcha_2014}. This participatory approach enables broader area coverage at finer spatial resolution. Methods have included human-carried smartphone sensors, wearable devices, and the use of big data from Google Street View cars \cite{xu_gotcha_2014, xu_gotcha_2016, liu_third-eye_2018, dutta_common_2009, hasenfratz_participatory_2012, apte_high-resolution_2017, wang_efficient_2017}.

However, vehicular and mobile sensing face challenges such as spatial gaps and route bias, as sensor-equipped vehicles do not follow uniform paths, leading to uneven data collection \cite{xu_incentivizing_2019, chen_pas_2020}. This limitation is particularly relevant to homelessness research, as encampment areas may be underrepresented in datasets if rideshare and taxi fleets avoid them. To mitigate these gaps, researchers have applied machine learning-based interpolation methods, physics-based models, and matrix completion techniques to estimate pollution levels in unmeasured areas and improve overall accuracy \cite{chen_adaptive_2022, chen_hap_2016, chen_pga_2018, liu_third-eye_2018, wang_efficient_2017, du_effective_2015}.

In the context of homelessness research, crowdsourced air quality sensing enables mapping of pollution exposure in areas with known homeless populations. Aggregating data from vehicles operating in different neighborhoods allows for the identification of persistent pollution hotspots near encampments. This information can guide public health interventions, such as expanding green spaces near shelters or installing air filtration systems in transitional housing. Furthermore, real-time pollution mapping may help predict encampment migration patterns, assisting city officials in allocating resources to high-risk areas proactively.

\subsection{Sampling Techniques and Future Directions}

Sampling techniques vary in spatial and temporal resolution, with different methodologies applied to static and mobile deployments. City-wide air quality sampling increasingly relies on mobile sensor networks to enhance spatial coverage and resolution. One effective approach involves equipping fleet vehicles with pollution sensors to collect high-resolution environmental data. For example, a study in Oakland, CA, used Google Street View cars to measure air pollution across a 30 km\textsuperscript{2} area, covering 750 road-km over a year. These vehicles sampled pollutants such as NO, NO\textsubscript{2}, and black carbon at a fine 30 m spatial resolution, revealing significant pollution variability within individual city blocks \cite{apte_high-resolution_2017}.

To improve efficiency, adaptive sampling strategies dynamically adjust data collection frequency based on pollutant variability. When pollution levels remain stable, the system reduces the sampling rate; in highly variable regions, it increases the frequency to ensure accuracy \cite{wang_efficient_2017, chen_adaptive_2022}. Additionally, grid-based monitoring divides urban areas into sections, ensuring comprehensive data collection across each cell. Traffic conditions, vehicle speed, and road topology influence sampling rates, optimizing resources while maintaining accuracy \cite{wang_efficient_2017, xu_ilocus_2020}. These strategies enhance pollution mapping and support interventions in real-time traffic management and public health.

However, passive adaptation cannot cover areas mobile sensors do not reach. This creates sensor holes and potential sampling bias. To address this, researchers have studied actuating vehicles or incentivizing paths that visit underrepresented areas~\cite{lin_neural-based_2022, joe-wong_taxi-for-all_2021, xu_incentivizing_2019, xu_vehicle_2019, chen_asc_2019}. Others have focused on optimizing sampling distribution across a region~\cite{wu_generative_2020, lin_neural-based_2022}. These approaches are especially relevant for homelessness research, as encampments often reside in low-traffic or less desirable areas.

Even with adaptive strategies, some regions may still lack adequate samples. To address this, machine learning and physics-informed models have been used to extrapolate pollution data from nearby samples and environmental features~\cite{ma2019deep, ma_fine-grained_2020, ma_enhancing_2020}.

By combining city-wide sensor networks, adaptive sampling, and predictive modeling, urban planners and policymakers can gain a more nuanced understanding of how air quality disparities affect homeless populations. These insights can guide targeted environmental interventions—such as pollution reduction near shelters or zoning policies to minimize exposure for vulnerable groups. Future research should integrate air quality data with real-time pedestrian movement, socioeconomic factors, and housing availability to build a holistic framework for understanding the intersection of environmental health and homelessness.
 
\section{Utilizing Internet of Things and Edge Computing}
The Internet of Things (IoT) and distributed computing in smart cities are well positioned to collect large-scale data efficiently. When evaluating urban conditions, distributed computing nodes can gather localized, fine-grained data at a low-cost. These nodes are often equipped with specialized sensors that collect various types of data. By combining multiple sensors at a single node, correlations between different urban factors can be examined in real-time \cite{cornelius_efficient_2020}.

Air quality and population density tracking are particularly well suited for IoT-based analysis \cite{ghoneim_towards_2019}. Prior studies have explored using GPS to map pollution at the neighborhood level, providing residents with relevant and personalized environmental insights. Since air quality can vary significantly even within a few city blocks, dense distributions of IoT nodes enable precise, localized measurements \cite{banach_new_2020, egodagama_air_2021, munera_iot-based_2021}. Embedded systems can also integrate sensors tailored to specific pollutants, such as CO\textsubscript{2} and O\textsubscript{3} levels \cite{siregar_integrated_2016}. While air pollution monitoring is a common application, IoT nodes have also been deployed for water and waste pollution tracking, demonstrating their versatility \cite{salman_review_2023}. 

A key advantage of embedded computing and IoT is their ability to perform on-device processing for efficient local data analysis. In particular, digital signal processing (DSP) techniques have been employed to filter and classify sources of pollution. This is especially useful for noise pollution analysis, where microcontrollers running lightweight DSP algorithms can effectively process sound data \cite{liu_internet_2020, lopez_digital_2020}. More advanced microcontroller-based systems leverage artificial neural networks (ANNs) or other machine learning algorithms to make intelligent predictions based on collected data \cite{banach_new_2020, asha_iot_2022}. Distributed nodes allow for processed, real-time data - allowing residents accurate and updated data about their neighborhoods \cite{badii_real-time_2020}.

Despite these benefits, deploying microcontroller-based sensor nodes at scale presents several challenges. Power consumption, energy efficiency, and network communication remain critical concerns. Many proposed solutions rely on battery or solar-powered systems to minimize maintenance and environmental impact \cite{balid_development_2017}. However, low-power constraints often conflict with the computational demands of on-device processing. For instance, ANN-based processing can require custom ASIC designs optimized for power efficiency while still delivering adequate performance.

Network infrastructure is another challenge. Reliable transmissions from distributed nodes is a difficult task to achieve. Many architectures employ a hierarchical tree structure, where local nodes communicate with access points that connect to the internet via cellular networks. Some systems directly interface with cellular networks, while others utilize WLAN or other localized wireless connections \cite{alahi_integration_2023, kanellopoulos_networking_2023}. Scaling these networks is complex; a high-volume of nodes transmitting data can overwhelm existing infrastructure \cite{jin_information_2014, tonjes2014real}. To prevent failures, proper Quality of Service (QoS) management is essential to enforce bandwidth limitations.

Finally, mass deployment of interconnected IoT nodes raises privacy and security concerns \cite{toma_iot_2019}. Microcontroller-based networking is typically designed to be as low-bandwidth as possible, limiting the feasibility of advanced cryptographic protocols. Research efforts have focused on-developing lightweight encryption algorithms that balance security, performance, and power efficiency. Addressing these privacy challenges will be crucial for the widespread adoption of IoT-based environmental monitoring systems.
\section{Assessing Neighborhoods Through Behavioral Monitoring}

Individuals' interactions with both the physical environment and each other significantly influence their movement patterns within a neighborhood. Recent studies utilizing eye-tracking and computer vision (CV) have explored how people experience the built environment, examining their perceptions of various features and how biases shape their interpretations. 











\subsection{Eye-Tracking for Environmental Perception}
With the increasing availability of eye-tracking technology, a body of work has shown that it is viable to associate subjects’ attention, as measured via eye-tracking \cite{piao_capturing_2013}, to how they experience a built environment (urban or not) \cite{noland_eye-tracking_2017, brazil_using_2017, tavakoli_psycho-behavioral_2025, hollander_seeing_2019, hakiminejad_shaping_2025}. While most of these studies were conducted with static images, in-situ live tracking on a mobile head-mounted platform was performed by Simpson et al., where researchers determined what aspects of the built environment people engage with most \cite{simpson_visual_2019}. 

Static images may encode researcher biases through the way the researchers select images \cite{noland_eye-tracking_2017}. Though eye-tracking gives more insight into what subjects fixate on within an image, static images chosen by the researchers still carry this potential bias; therefore, removing confounding variables by placing the subjects in an authentic environment is most preferable. 
As mentioned in Simpson et al. \cite{simpson_visual_2019}, eye-tracking still runs into issues with data loss due to bright light and the user's movements \cite{evans_collecting_2012, tomasi_mobile_2016}, and the fact that attention not being fully correlated with gaze is exacerbated in a stimulating environment like a busy city street \cite{uttley_eye-tracking_1}. 

Past work with walking interviews also demonstrates some benefits by having subjects interact with a real environment. Lauwers et al. mentions that a strength of the walking interview lies in that interviewer/interviewee can be fully immersed in the environment, with disruptions and detours that have the chance to add insight into conversations \cite{lauwers_exploring_2021}. Similarly, we hypothesize that immersion within a real environment carries similar benefits, in that researchers can gain more insights through natural interruptions and disturbances that the subject experiences. 






\subsubsection{CV from Mobile Platforms}
Tracking from a mobile platform poses difficulties across sensor modes, as assumptions based on static environments often fail. This requires moving away from simpler pre-trained object detectors (tracking by detection) and towards distinguishing 3D objects, then assessing how they move (tracking-before-detection) \cite{baumgartner2013tracking}. 






\subsection{Interpreting Pedestrian Behavior}

Though much of the work in identifying behavior of people moving through an area has not been extended across multiple modalities of transport, pedestrian behavior is well characterized. 

Tracking isolated pedestrians is relatively simple. 
However, problems arise with more people entering the field-of-view of the sensor because the presence of more people increases the potential for data loss via occlusion. To address this, we can identify group behaviors to mitigate the discrepancies in models \cite{hutchison_unified_2012, lau_tracking_2009} and simplify tracking by tracking groups of pedestrians rather than individuals. 
Improving our models of pedestrian trajectories can lessen the impact of occlusion. Improving on initial assumptions of Brownian (random) motion, it can be assumed that pedestrians have a destination and will efficiently path towards it; therefore, identifying potential destinations for pedestrians can improve accuracy during occlusion by predicting motion \cite{bruce_better_2004}. Multiple studies have unified these two concepts \cite{hutchison_unified_2012, yamaguchi_who_2011, pellegrini_improving_2010} and additionally demonstrate unspoken social behaviors related to environmental goals and pedestrian trajectories, which can be exploited to simplify pedestrian tracking further. Behaviors may also become more predictable by assessing the objects that move with a pedestrian \cite{baumgartner2013tracking}. 

For tracking higher-density crowds using CV, where pretrained object detectors and facial recognition methods fail due to top-down perspective, occlusion, and low resolution, work has been done to use the crowd itself to speed up computation of multi-target tracking for hundreds of people \cite{dehghan_binary_2018}.

Much of the recent work in crowd-tracking via CV methods—using static and dynamic platforms from ground-based and aerial views across varying crowd densities—is covered in a survey by Chaudhry et al. \cite{chaudhry_crowd_2019}.

Despite most of these studies being conducted using computer vision as the primary means of sensing, Lau et al. \cite{lau_tracking_2009} demonstrated that range data taken from a mobile platform can also be used to identify groups of people, which lends promise to relying on other ranging techniques to do the same. 








\section*{Conclusion}

The integration of environmental sensing, machine learning, and IoT technologies offers a scalable, data-driven approach to addressing the growing homelessness crisis. Recent research demonstrates the feasibility of using computer vision for encampment detection, air quality sensing for environmental risk assessment, IoT and edge computing for distributed data collection, and pedestrian tracking for behavioral analysis in urban spaces. While these advancements significantly improve upon traditional manual methods, several challenges remain.

First, mobile and crowdsourced sensing deployments must balance the need for broad coverage with cost constraints. Sparse and inconsistent data necessitate robust fusion techniques to integrate multisource environmental and visual data into high-resolution insights. Second, real-time processing is constrained by computational limitations, particularly for deep learning models running on resource-limited edge devices. Optimizing these models for efficiency without sacrificing accuracy is critical for practical deployment. Third, data privacy and ethical considerations must be carefully addressed, particularly when tracking movement patterns or analyzing sensitive information. Ensuring responsible data collection and compliance with privacy regulations is essential for public trust and adoption. Future research should explore adaptive sensing techniques capable of dynamically adjusting to shifts in population distribution and environmental conditions. Collaboration among researchers, city officials, and social organizations is crucial to translating sensing-driven insights into actionable policies and targeted resource allocation strategies.

By addressing these research gaps and technical challenges, future advancements in environmental sensing and computational methodologies can enhance the detection of homelessness, improve public health interventions, and contribute to more effective, data-driven urban policy planning.

\renewcommand{\bibfont}{\scriptsize}

\bibliographystyle{ACM-Reference-Format}
\bibliography{references.bib}

\end{document}